\documentclass[prl,twocolumn,a4paper,amsmath,floatfix]{revtex4-1}
\usepackage[utf8]{inputenc}
\usepackage{graphicx}
\usepackage{amsmath}

\usepackage{color}
\begin{document}


\title{Inverse design of soft materials via a deep-learning-based evolutionary strategy}




\author{Gabriele M. Coli}
\affiliation{Soft Condensed Matter, Debye Institute for  Nanomaterials Science, Utrecht University, Princetonplein 1, 3584 CC Utrecht, Netherlands}

\author{Emanuele Boattini}
\affiliation{Soft Condensed Matter, Debye Institute for  Nanomaterials Science, Utrecht University, Princetonplein 1, 3584 CC Utrecht, Netherlands}

\author{Laura Filion}
\affiliation{Soft Condensed Matter, Debye Institute for  Nanomaterials Science, Utrecht University, Princetonplein 1, 3584 CC  Utrecht, Netherlands}

\author{Marjolein Dijkstra}
\affiliation{Soft Condensed Matter, Debye Institute for Nanomaterials Science, Utrecht University, Princetonplein 1, 3584 CC  Utrecht, Netherlands}
 


\begin{abstract}
Colloidal self-assembly -- the spontaneous organization of colloids into ordered structures -- has been considered key to produce next-generation materials. However, the present-day staggering  variety of colloidal building blocks and the limitless number of thermodynamic conditions make a systematic exploration intractable. The true challenge in this field is to turn this logic around, and to develop a robust, versatile algorithm to inverse design colloids that self-assemble into a target structure.
Here, we introduce a generic inverse design method to efficiently reverse-engineer crystals, quasicrystals, and liquid crystals by targeting their diffraction patterns. 
Our algorithm relies on the synergetic use of an evolutionary strategy for parameter optimization,  and a convolutional neural network as an order parameter, and provides a new way forward for the inverse design of experimentally feasible colloidal interactions, specifically optimized to stabilize the desired structure.
\end{abstract}

\maketitle

\section{Introduction}

\begin{figure}[!h]
\centering
\includegraphics[width=0.45\textwidth]{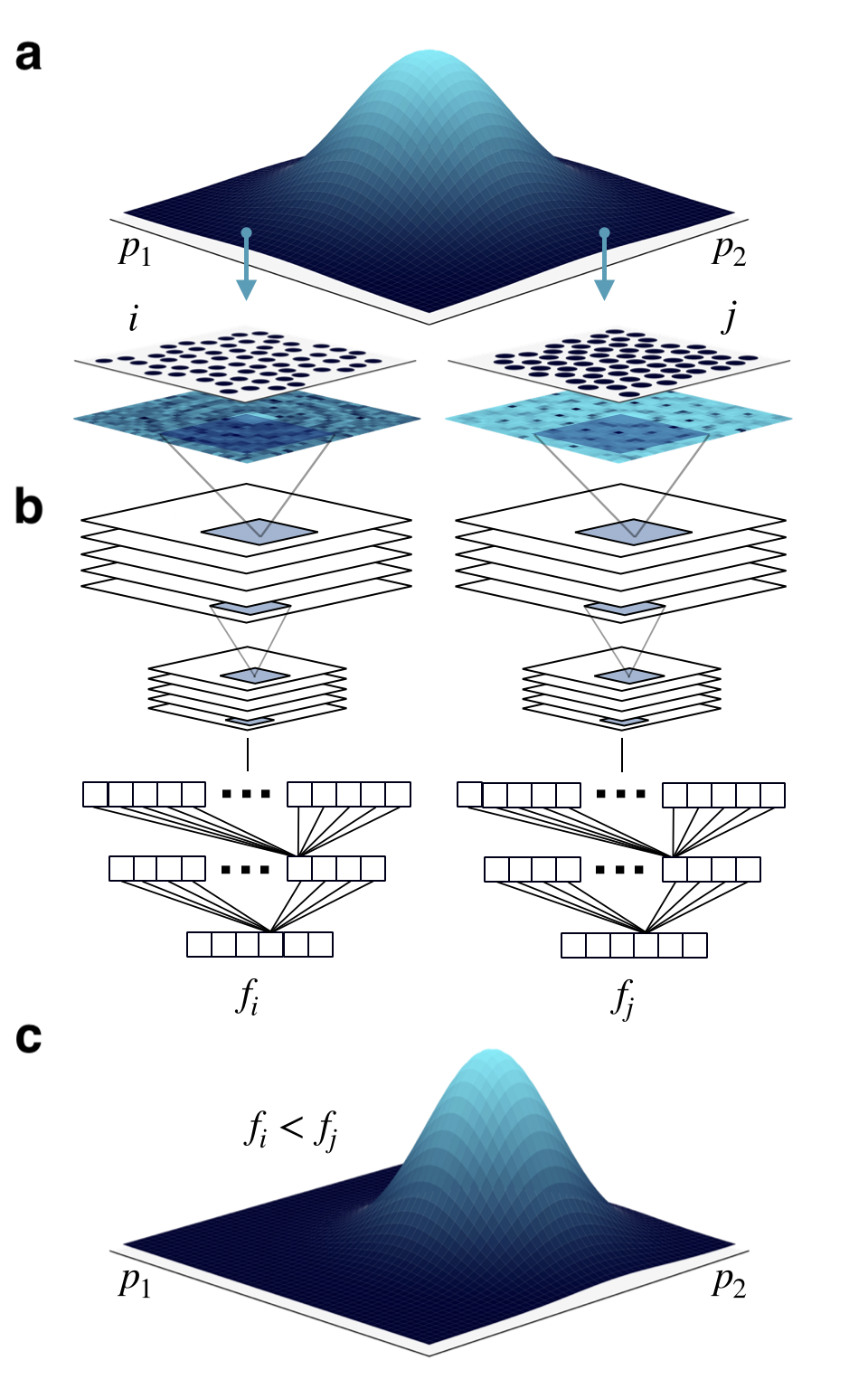}
\caption{{\bf Schematic representation of the three steps performed at each generation.} (a) In the first step, we draw candidate sets of parameters ($p_1$ and $p_2$ in the figure) from a multivariate Gaussian distribution. For each set, or sample, we then perform a simulation. (b) In the second step, samples are ranked and scored based on their fitness $f$, which is evaluated using a convolutional neural network trained to classify phases according to their diffraction patterns. Samples with a higher likelihood of being classified as the target phase will be scored with a higher fitness. (c) In the third and final step, the Gaussian distribution is updated in order to move towards regions of the parameter space where the fittest samples have been encountered.}
\label{fig:method}
\end{figure}

Self-assembly of colloidal particles is ubiquitous in nature and is considered to be of paramount importance for the design of novel functional materials. For example, viruses, lipid bilayers, tissues, atomic and molecular crystals, liquid crystals, and nanoparticle superlattices are all self-assembled from smaller components in a highly intricate way. The structure of such an assembly is determined by the interactions of the building blocks and by the thermodynamic conditions, e.g. pressure, temperature, or composition.  Understanding the relation between building blocks and self-assembled arrangements is essential for materials design as the physical properties of materials are intimately related to the structure. 

On the other hand, huge progress has been made over the past decades in the synthesis and fabrication of colloidal particles, resulting in a spectacular variety of novel colloidal building blocks to the point where particles with any shape and interaction potential can be made on demand \cite{glotzer2007anisotropy,sacanna2011shape,miszta2011hierarchical,boles2016self,he2020colloidal}. Traditionally, tremendous efforts have been devoted to the “forward design” problem: Which structures with what properties are formed for a given colloidal building block under what circumstances? A major drawback of this approach is that the number of possible building blocks and thermodynamic conditions is limitless, making a systematic exploration of these design spaces intractable.

The true challenge in materials science is to develop a robust, versatile algorithm for solving the “inverse design” problem and to design building blocks that self-assemble into a target structure. The lack of such an inverse design method (IDM) forms a significant obstacle for the full exploitation of colloidal self-assembly in the development of tomorrow’s materials \cite{rechtsman2005optimized,florescu2009designer,miskin2013adapting,jain2014inverse}.

In this work, we present a general inverse design method based on deep-learning techniques to reverse engineer a multitude of thermodynamic phases, ranging from crystals, to liquid crystals and even quasicrystals. A novel machine-learning-based order parameter is combined with an evolutionary strategy which searches the multi-dimensional parameter space to optimize the colloidal interactions and thermodynamic conditions (density, temperature, etc.) for the self-assembly of a target phase. 

Designing an IDM to reverse engineer phases, from crystals, to liquid crystals,  and quasicrystals, generally requires two ingredients. First, one should define an   order parameter that is sensitive to the global structure of a multitude  of phases and  can be exploited as a fitness function indicating how ``close'' one is to the desired outcome. Secondly, one has to devise a mathematical scheme to update  the design parameters based on the chosen fitness function.

The latter requirement can be easily satisfied by choosing among several techniques, either borrowed from classical optimization algorithms   \cite{long2018rational,kumar2019inverse,khadilkar2017inverse} or inspired by statistical physics \cite{miskin2016turning,geng2019engineering,kumar2019inverse}. Our inverse design method (IDM) uses the Covariance Matrix Adaptation (CMA) evolutionary strategy for parameter optimization \cite{hansen2006towards,kumar2019inverse}. 

Conversely, the choice of an effective fitness function represents the real bottleneck for any IDM to succeed. In the last decade, a plethora of order parameters has been used to define fitness functions for all kinds of phases. 
For instance, free-energy or chemical-potential differences with respect to the competing structures have been employed to reverse engineer 3D crystal lattices starting from (non)spherical colloids \cite{jain2013inverse,van2015digital}. Often,  full knowledge of the target crystal has been translated into a fitness function by computing the mean square displacements of the particles with respect to their target lattice points \cite{rechtsman2005optimized}, or through the radial distribution function \cite{jadrich2015equilibrium,pineros2018inverse,sherman2020inverse}. The sometimes unrealistic resulting potentials have been explicitly filtered by Adorf \emph{et al.} in order to obtain smooth and short-range interactions \cite{adorf2018inverse}.

Although all these fitness function definitions brilliantly achieve their goals, they often lack generality,  and most importantly, they are not able to simultaneously and equally penalize competing phases. In other words, they do not have the ability to create an approximately flat fitness landscape, where the design engine can move smoothly, with only one preferred region corresponding to the target phase. Moreover, in the case of quasicrystals, in spite of the certified need of two inherent length scales in the system \cite{barkan2011stability,dotera2014mosaic}, the actual positions of the constituent particles remain unknown, therefore representing a significant challenge to the above strategies.

Inspired by the highly successful history of identifying phases by their scattering patterns in combination with advances in machine learning (ML), we attack the problem from a new avenue and directly use an encoding of the structure factor as the order parameter. To this end, we train a convolutional neural network (CNN) to classify different phases from their diffraction pattern, and use the result to construct a fitness function, such that configurations with a higher likelihood of being classified as the target phase, will be scored with a higher fitness.
A sketch of the final algorithm is shown in Fig. \ref{fig:method}.

This algorithm turns out to be extremely robust and versatile, facilitating the inverse design of, not only crystal and liquid crystalline phases, but also quasicrystals -- which due to their non-periodicity are notoriously difficult to inverse design.

\section{Results}

\noindent \textbf{Our Inverse design method:}
Our IDM combines the CMA evolutionary strategy for parameters optimization, and a CNN for the fitness evaluation, which are both described in detail in the Methods section.  The goal is to optimize the free parameters of a given model in order to favour the formation of a target phase. 

The method proceeds in generations, or iterations, consisting of essentially three steps: a) sampling, b) fitness evaluation, and c) update. In the following, we give a general overview of these three steps, which are sketched in Fig. \ref{fig:method}.

In the first step (Fig. \ref{fig:method}a), we draw a fixed number of candidate sets of parameters from a multivariate Gaussian distribution. The dimension of this multivariate Gaussian distribution is determined by the number of design parameters that we wish to tune. For each candidate set of parameters, we then perform a simulation of the system and save a number of representative configurations.
In the second step (Fig. \ref{fig:method}b), we score and rank the samples based on their fitness $f$. In general, the fitness is a measure of similarity between a sample and a specific target, and it is maximized when the target is reached. 
Here, we introduce a new fitness function based on CNNs that are trained to classify different phases based on their diffraction patterns. We use this CNN to process the configurations saved during each simulation, and assign a larger fitness to samples with a higher probability of being classified as the target phase.
Finally, based on this score, the mean and the covariance matrix of the multivariate Gaussian distribution are updated using the CMA equations,  which are designed  to facilitate an efficient exploration of parameter space. As sketched in Fig.  \ref{fig:method}c, the update not only allows the mean of the distribution to move towards regions with a higher fitness, but it also speeds up sampling by stretching the distribution when several updates are in the same direction, and then shrinking it once the fitness is maximized. This whole procedure is repeated multiple times until the fitness is maximized and/or a predetermined convergence criterion is met. \\

\noindent \textbf{Setting up the IDM in two dimensions:}
The first model we consider is a  two-dimensional system in which the particles interact with a hard-core square-shoulder (HCSS)  potential:
\begin{equation}
\beta u(r)= 
\begin{cases}
    \infty,& r < \sigma\\
    \beta\epsilon,              & \sigma \leq r \leq \delta \\
    0, & r > \delta,
\end{cases}
\label{hcss}
\end{equation}
with $r$ the center-of-mass distance between two particles, $\epsilon$ the interaction strength,  $\sigma$ the core diameter, $\delta$ the  interaction range, and $\beta = 1/k_B T$ with $k_B$ Boltzmann's constant and $T$ the temperature.
This model has been shown to self-assemble into a variety of phases \cite{dotera2014mosaic,pattabhiraman2015stability,pattabhiraman2017phase,pattabhiraman2017formation}, including several crystal structures and various  quasicrystals (QCs), which makes it an ideal playground for setting up and testing our IDM. The three QCs we consider here, which are the dodecagonal (QC12), the decagonal (QC10), and the octadecagonal (QC18) quasicrystals, are found to be stable for different values of the interaction range $\delta$, and only in a tiny range of densities $\rho$ and temperatures $T$. In all cases we explore, the competing stable phases include the fluid, the hexagonal (HEX) crystal, and the square (SQ) crystal phase. 

To set up our IDM we trained a CNN to classify the aforementioned phases based on their two-dimensional diffraction patterns, as described in the Methods. Specifically, the CNN takes as input the diffraction pattern of a given configuration, and outputs a vector of real numbers with as many components as the number of phases to distinguish. Each number in the output is indicative of the likelihood that the given input corresponds to one of the phases. This output is then used to define the fitness function to target a specific phase. 

The data set for training the CNN is built by performing MC simulations of the HCSS model in the $NPT$ ensemble. For each phase, we perform simulations at different state points, and collect a large number of independent configurations. The set of diffraction patterns generated from these configurations constitutes the data set on which the CNN is trained and validated. Overall, we find the CNN to be highly effective and able to classify all phases with $100\%$ accuracy. \\

\begin{figure*}
\centering
\includegraphics[width=1.0\textwidth]{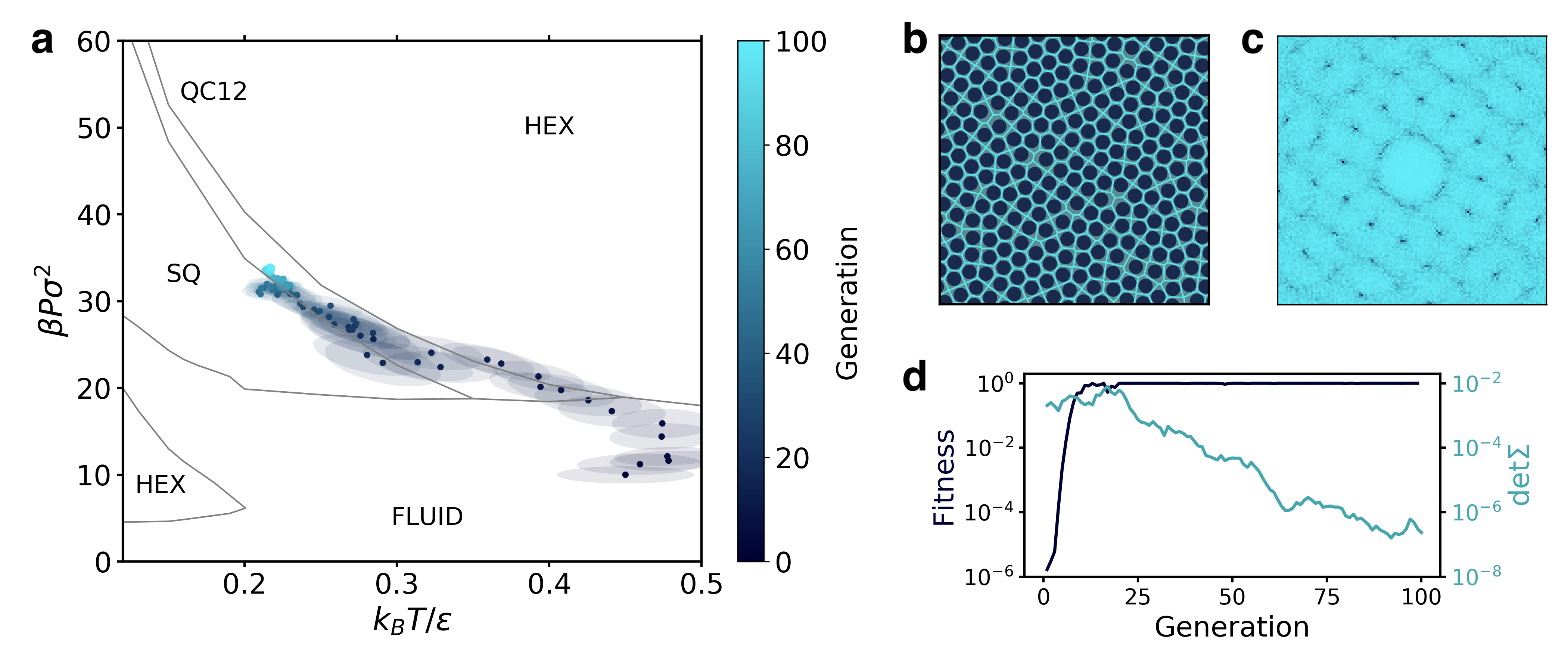}
\caption{{\bf Reverse engineering of the QC12 in the HCSS model.} (a) Evolution of the Gaussian distribution in the $k_BT/\epsilon-\beta P\sigma^2$ plane. Points and ellipses represent the mean and the covariance matrix (within one standard deviation) of the distribution. The phase diagram in the background is adapted from Ref. \onlinecite{pattabhiraman2015stability}. (b) Representative snapshot of the QC12 obtained during the last generation. The hard cores are shown in a dark color, while lines show their Voronoi tessellation. (c) Diffraction pattern of the snapshot in (b). (d) Evolution of the mean fitness and the determinant of the covariance matrix.}
\label{fig:trajfluid}
\end{figure*}

\noindent \textbf{Reverse engineering of the QC12 in the HCSS model:} We start our investigation by considering the HCSS model with a fixed value of the shoulder width $\delta=1.4\sigma$, at which the QC12 phase has been shown to be stable \cite{pattabhiraman2015stability,pattabhiraman2017phase}. The phase diagram as a function of temperature and pressure (adapted from Ref. \cite{pattabhiraman2015stability}) is reported in Fig. \ref{fig:trajfluid}a.

The goal here is to reverse engineer the QC12 phase by letting the evolutionary strategy find the narrow region in the phase diagram  where the QC12 phase is stable by tuning the system parameters  pressure $P$ and temperature $T$. In other words, we keep the interaction parameters fixed, while trying to optimize the thermodynamics variables to favour the formation of the QC12. Our knowledge of the phase diagram allows us to easily asses and monitor the performance of the reverse engineering process.

To explicitly target the QC12, we use the output of the trained CNN to define the fitness function $f$ for the evolutionary strategy. In particular, for any sample, i.e. for any simulation, we define the fitness as $f=\bar{P}_{\text{QC12}}$, where $P_{\text{QC12}}$ is the probability that the diffraction pattern of a given configuration is classified as a QC12 by the CNN, and the bar indicates an average taken over representative configurations visited during the simulation.

The results of the reverse engineering process are summarized in Fig. \ref{fig:trajfluid}. Starting the reverse engineering process with a Gaussian centered in the region of stability of the fluid phase, the algorithm  reaches the region where the target QC12 is stable in approximately $25$ generations. Fig. \ref{fig:trajfluid}a shows the evolution of the multivariate Gaussian distribution in the temperature $k_BT/\epsilon$-pressure $\beta P \sigma^2$ plane across successive generations.
A representative snapshot obtained in the last ($100$th) generation is shown in Fig. \ref{fig:trajfluid}b, while the corresponding diffraction pattern, characterized by twelve-fold rotational symmetry, is shown in Fig.\ref{fig:trajfluid}c.

The success of the algorithm heavily relies on the ability of the CNN to spot even small structural variations in the system. At the early stages of the reverse engineering process, when the system is in the fluid phase, the algorithm already finds it convenient to increase the pressure, and hence the density, in order to increase the overall structural order. This can clearly be seen in Fig. \ref{fig:trajfluid}d, where we plot the evolution of the mean fitness averaged over all samples. Although the variations of the fitness in the early generations are very tiny, they are sufficient to guide the evolutionary strategy in the right direction.

An efficient exploration of phase space is then made possible by the CMA equations, which evolve the Gaussian distribution at each generation. This not only allows the mean of the distribution to move towards regions with a higher fitness, but it also allows the covariance to stretch when several updates are in the same direction, and then shrink once the fitness is maximized. This is shown in Fig. \ref{fig:trajfluid}d, where we plot the evolution of both the mean fitness and the determinant of the covariance matrix. The determinant becomes larger when the fitness improves, and it decays exponentially once the fitness is maximized.

Note that here we initialized the mean of the Gaussian distribution at a specific state point within the region of stability of the fluid phase, but we find the algorithm to be largely robust to changes in the initial conditions. In the SI, we show additional trajectories of the reverse engineering of the QC12 obtained by starting with a Gaussian distribution centered at different state points, i.e. in the fluid phase, the SQ phase, the HEX phase at relatively high temperature and low pressure, and the HEX phase at relatively low temperature and high pressure. In all cases, the mean of the parameters distribution converges to the region of stability of the target QC12, clearly showing that the performance is not affected by the particular choice made for the initial conditions.

Furthermore, we would like to stress a crucial aspect that demonstrates the versatility of the algorithm. In fact, the same method, and the exact same CNN, can be used to target any phase that was included in the training data set, simply by changing the definition of the fitness. For instance, to reverse engineer the hexagonal crystal phase, it is sufficient to impose $f = \bar{P}_{\text{HEX}}$. A trajectory of the reverse engineering of the hexagonal crystal is shown in the SI.\\

\begin{figure*}
\centering
\includegraphics[width=1.0\textwidth]{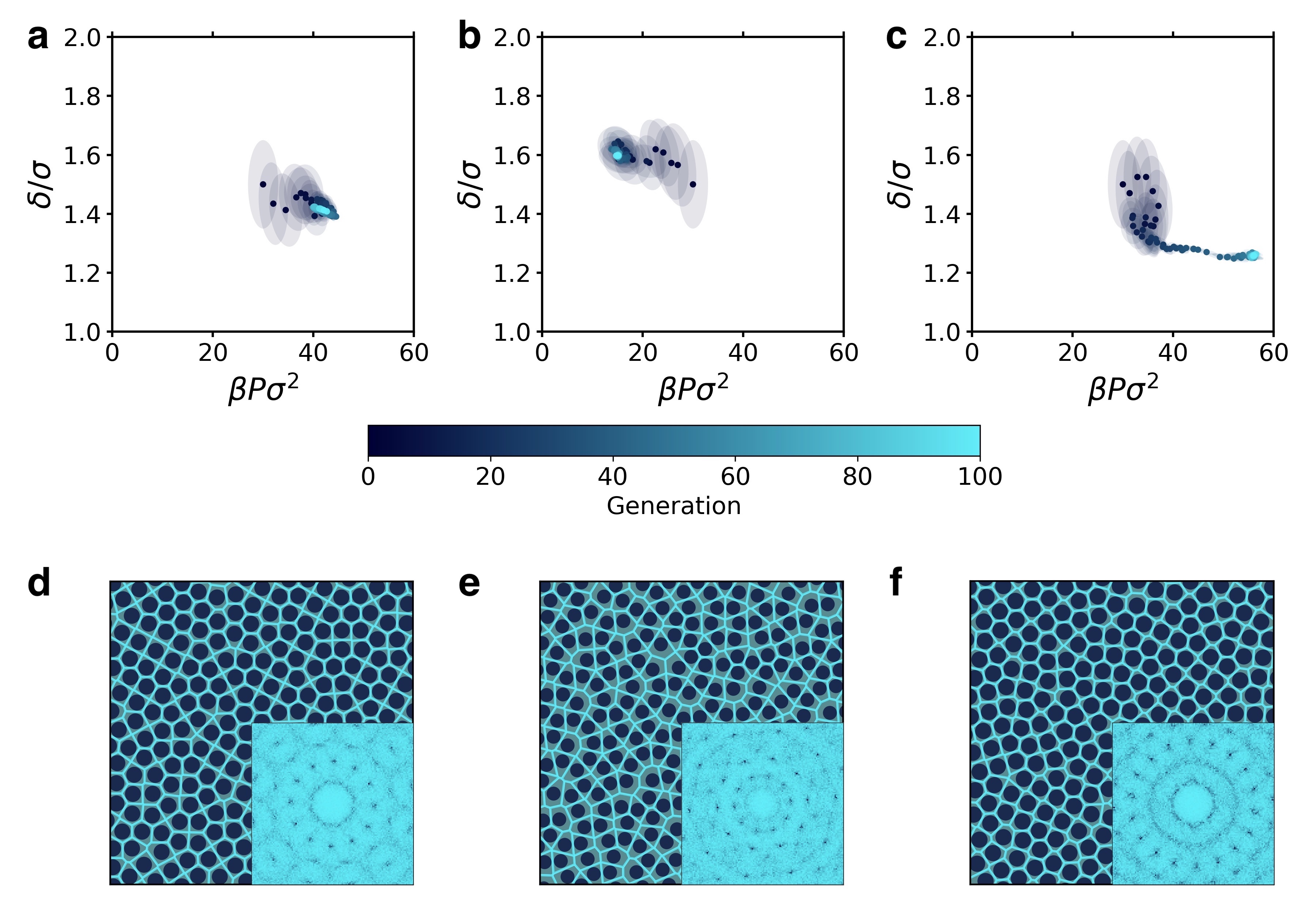}
\caption{{\bf Reverse engineering of QC12, QC10, and QC18 in the HCSS model.} (a-c) Evolution of the Gaussian distribution in the $\beta P\sigma^2-\delta/\sigma$ plane during the reverse engineering of (a) the QC12, (b) the QC10, and (c) the QC18 phases. Points and ellipses represent the mean and the covariance matrix (within one standard deviation) of the distribution. (d-f) Representative snapshots of the (d) QC12, (e) QC10, and (f) QC18 obtained in the last generation, along with their diffraction patterns and Voronoi tessellations.}
\label{fig:allqcs}
\end{figure*}

\noindent \textbf{Reverse engineering of QC12, QC10, and QC18 in the HCSS model:} As already discussed, in addition to the QC12, the HCSS model exhibits two other quasicrystalline structures, which are stabilized for different values of the shoulder width $\delta$. As a natural next test, we now explore whether we can reverse engineer all the three stable quasicrystals (QC12, QC10, and QC12) considered in this work. To this end, we fix the temperature to $k_BT/\epsilon = 0.17$, a temperature for which all three QCs are stable, and let the evolutionary strategy optimize the shoulder width $\delta$ and the pressure $P$ for each specific QC. In all three cases, we start the reverse engineering process from the same state point in the fluid phase ($\delta=1.5\sigma$ and $\beta P\sigma^2=30$), and choose the fitness function appropriate for the target phase.
The results of the reverse engineering process are summarized in Fig. $\ref{fig:allqcs}$. In particular, Figs. \ref{fig:allqcs}a-c show the evolution of the multivariate Gaussian distribution when targeting (a) the QC12, (b) the QC10, and (c) the QC18. Depending on the QC to be found, the distribution evolves in different directions, and eventually converges to different state points. In all cases, the final values of pressure and shoulder width obtained are in excellent agreement with those at which the three QCs have been shown to be stable \cite{dotera2014mosaic,pattabhiraman2015stability,pattabhiraman2017phase,pattabhiraman2017formation}. Representative snapshots of the QCs obtained and their diffraction patterns are shown in Fig. \ref{fig:allqcs}d-f. Each diffraction pattern immediately confirms the presence of the correct quasicrystalline structure. \\

\begin{figure*}[!h]
\centering
\includegraphics[width=1.0\textwidth]{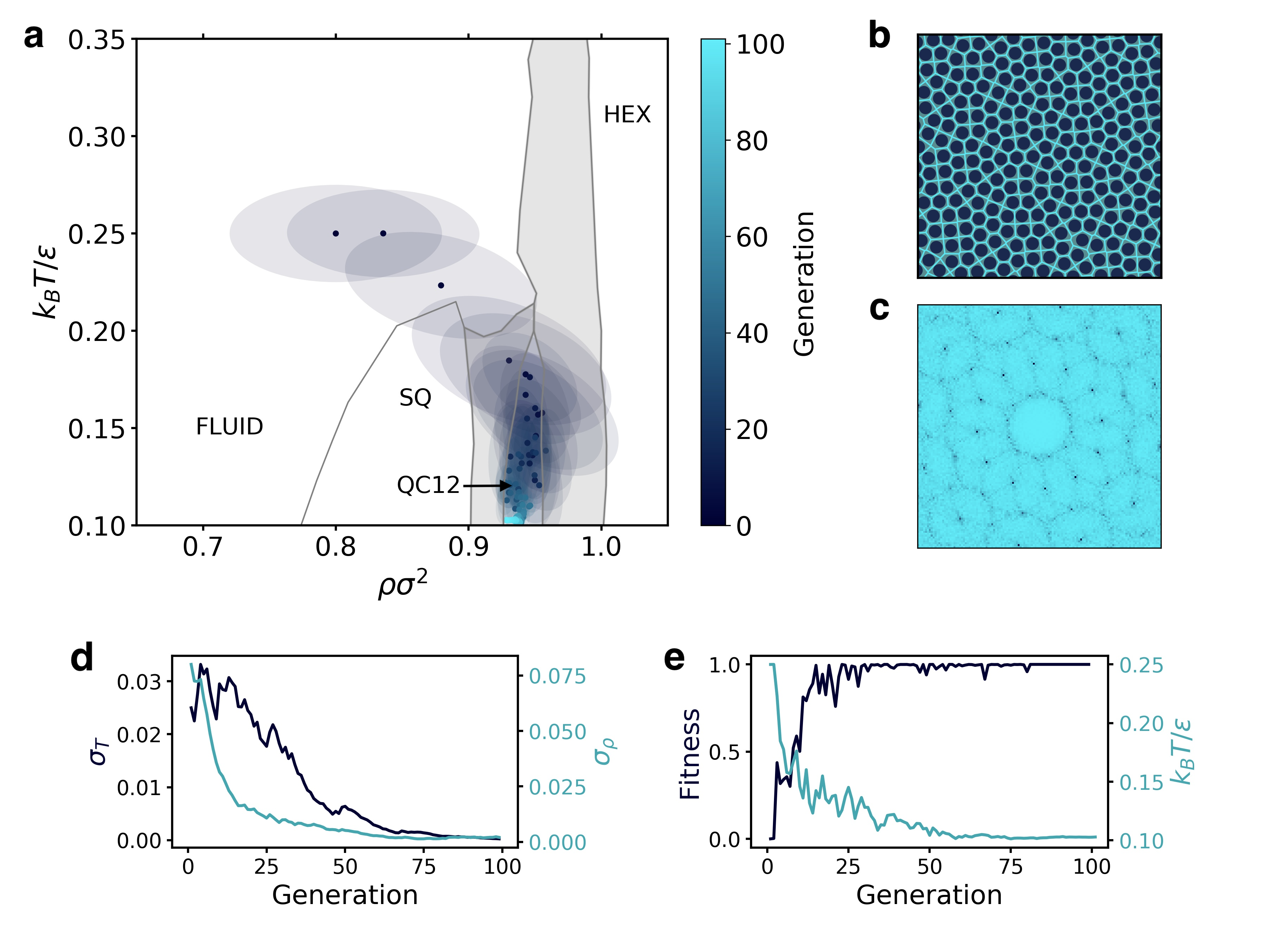}
\caption{{\bf Reverse engineering of the QC12 in the SCS model.} (a) Evolution of the Gaussian distribution in the $\rho\sigma^2-k_BT/\epsilon$ plane. Points and ellipses represent the mean and the covariance matrix (within one standard deviation) of the distribution. The phase diagram in the background is adapted from Ref. \onlinecite{padilla2020phase}. Coexistence regions are indicated in light gray. (b) Representative snapshot of the QC12 obtained during the last generation and its Voronoi tessellation. (c) Diffraction pattern of the snapshot in (b). (d) Evolution of the square root of the covariance matrix's diagonal elements, which correspond to the standard deviations along the temperature ($\sigma_T$) and density ($\sigma_\rho$) directions.  (e) Evolution of the mean fitness and the mean temperature in (a).}
\label{fig:softqc12}
\end{figure*}

\noindent \textbf{Application to a new model interaction:} Thus far we have only addressed the model that was used for training the CNN. A natural next question is whether the method is  general enough to work on other model systems, {\it without} having to retrain the CNN for the specific model under consideration.
To answer this question, we now consider a two-dimensional softened-core-shoulder (SCS) model with an interaction potential given by:
\begin{equation}
u(r)/\epsilon= \left( \frac{\sigma}{r} \right)^{14} + \frac{1 - \tanh [k(r-\delta)]}{2},
\label{eq:scs}
\end{equation}
where $\epsilon$ is the energy scale, $\sigma$ represents the typical core diameter, and $k$ and $\delta$ are two parameters that, respectively, control the steepness and the characteristic interaction range.
Similar to the HCSS, the QC12 has been shown to be stable in a limited range of densities and temperatures with a shoulder width of $\delta=1.35\sigma$ and $k\sigma=10$ \cite{kryuchkov2018complex, padilla2020phase}.

To test the ability of our method to be effective on new types of interactions, we use the same CNN that was trained on the HCSS model in order to reverse engineer the QC12 in the SCS model. Similar to the HCSS case, we keep the interaction parameters fixed, i.e. $\delta=1.35\sigma$ and $k\sigma=10$, and let the evolutionary strategy find the region of densities and temperatures in which the QC12 is stable. The phase diagram in Fig. \ref{fig:softqc12}a is used as a reference to asses and monitor the performance of the method. Note that, since this phase diagram is in terms of density and temperature, simulations are now performed in the canonical ensemble. Moreover, in contrast to the HCSS case, there are now stable coexistence regions between multiple phases (indicated with a gray background in Fig. \ref{fig:softqc12}a). As the CNN was not trained on configurations with a phase coexistence, this represents a further robustness test for our method.

The results of the reverse engineering process are summarized in Fig. \ref{fig:softqc12}. Specifically, Fig \ref{fig:softqc12}a shows the evolution of the multivariate Gaussian distribution in the temperature-density plane. Starting with a distribution centered in the fluid region, the algorithm immediately starts to increase the density and lower the temperature in order to increase the overall order. Impressively, after only 5 generations, the mean of the distribution is already inside the region of stability of the QC12, demonstrating the robustness of the CNN to changes in the interaction potential. In the remaining generations, the covariance of the distribution shrinks, and the mean moves towards lower temperatures in the phase diagram. A representative snapshot of the QC12 obtained during the last generation and its diffraction pattern are shown in Figs. \ref{fig:softqc12}b and \ref{fig:softqc12}c, respectively. 

Looking more closely at the evolution of the model parameters, it is interesting to observe the different behaviour of the temperature and density components. After the first 5 iterations, the density simply oscillates in the tiny range of stability of the QC12, while a large exploration keeps happening in temperature. This can be seen also by looking at the evolution of the standard deviations of temperature ($\sigma_T$) and density ($\sigma_{\rho}$) in Fig. \ref{fig:softqc12}d. While $\sigma_{\rho}$ decays almost monotonously from the very beginning, $\sigma_T$ oscillates for about 20 generations before starting its decay. 

We would also like to stress that the reason why the algorithm seems to prefer lower temperatures, despite being already in the stability region of the target phase, is mainly related to a decrease in the thermal fluctuations. With a lower amount of thermal noise, the CNN is presented with cleaner configurations, which, on average, are scored with a higher fitness. This can be seen in Fig. \ref{fig:softqc12}e, where we plot the evolution of both the mean fitness and the mean temperature. As the temperature goes down, the fluctuations of the fitness become smaller.\\

\begin{figure*}
\centering
\includegraphics[width=1.0\textwidth]{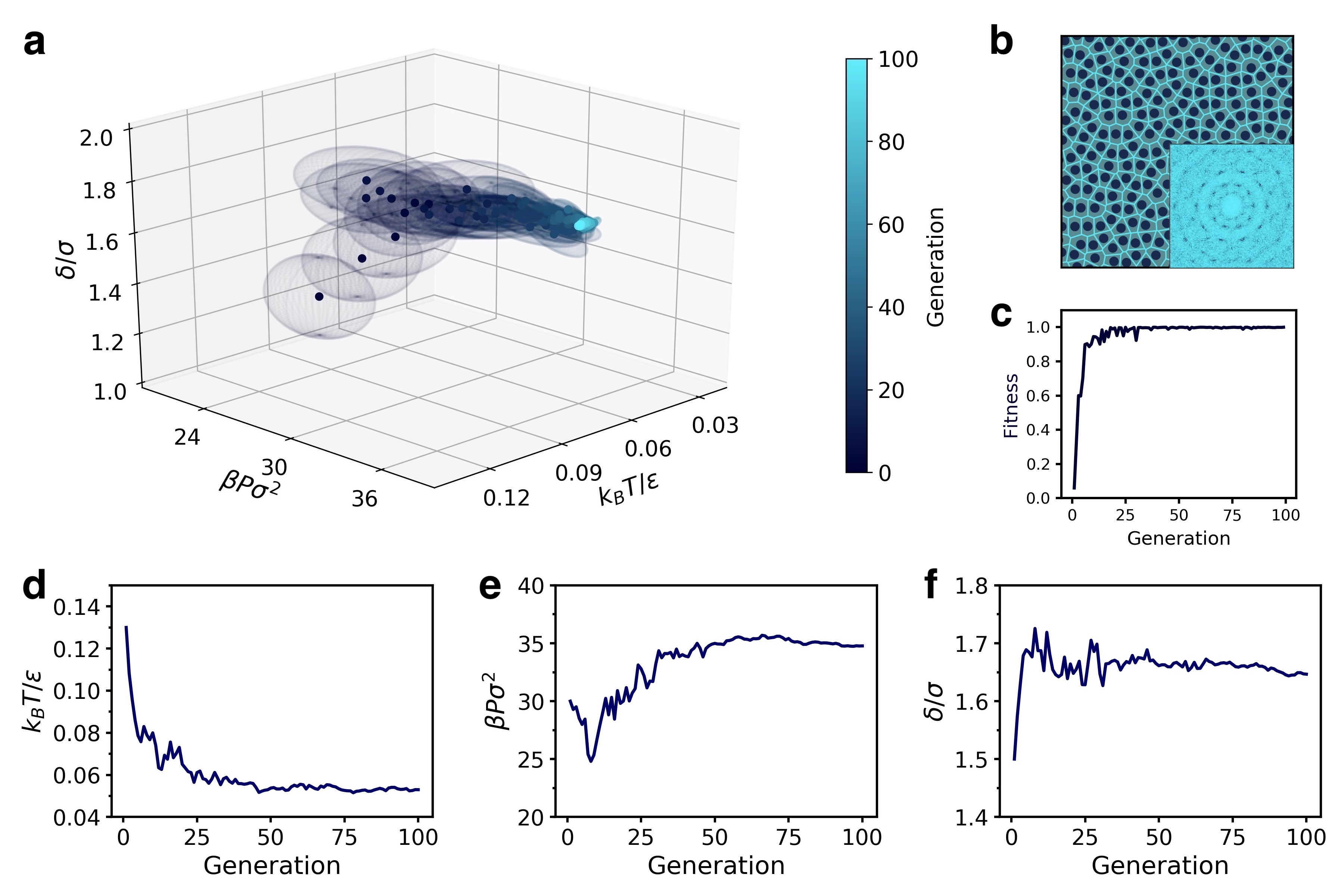}
\caption{{\bf Discovery of the QC10 in the SCS model.} (a) Evolution of the Gaussian distribution in $k_BT/\epsilon-\beta P\sigma^2-\delta/\sigma$ space. Points and ellipsoids represent the mean and the covariance matrix (within one standard deviation) of the distribution. (b) Representative snapshot of the QC10 obtained during the last generation, along with its diffraction pattern and Voronoi tessellation. (c) Evolution of the mean fitness. (d-f) Evolution of the three parameters in (a) optimized in the reverse engineering process: (d) temperature $k_BT/\epsilon$, (e) pressure $\beta P \sigma^2$, and (f) shoulder width $\delta/\sigma$.}
\label{fig:discovery10}
\end{figure*}

\noindent \textbf{Phase discovery:}
The fundamental ability of the algorithm to generalize to different interaction potentials opens up the possibility of discovering quasicystals in new model systems. For instance, given the similarities between the SCS and the HCSS models, we might ask whether also the SCS model stabilizes different quasicrystals for different values of the shoulder width $\delta$. We note that, compared to the HCSS model, much less is known about the phase behaviour of the two-dimensional SCS system.

Here, we explore the possibility of the SCS model to form a QC10. To this end, we fix $k\sigma=10$ as in the previous case, and let the evolutionary strategy optimize three parameters: shoulder width $\delta$, temperature $T$, and pressure $P$. Note that, by varying these three parameters simultaneously, the algorithm might encounter phases that were not included in the data set for training the CNN. We do not expect this to be a problem, as long as no phase is misclassified as the target phase. This could possibly cause the algorithm to get stuck and eventually converge to the wrong phase. While this problem did not occur in our test, a simple solution would be to include the newly found phase in the training data set, and retrain the CNN. 

The results of the reverse engineering process are summarized in Fig. \ref{fig:discovery10}. Starting from a fluid phase, the evolutionary strategy decreases the temperature, and  increases both the pressure and shoulder width in order to maximize the fitness (see Fig. \ref{fig:discovery10}c-f), discovering the not-yet-predicted QC10 phase for this system. As a further confirmation that the algorithm has indeed found a QC10, Fig. \ref{fig:discovery10}b shows a representative snapshot obtained during the last generation, along with the corresponding diffraction pattern. Hence, our algorithm has successfully located a new phase in the SCS model. \\

\begin{figure*}
\centering
\includegraphics[width=1.0\textwidth]{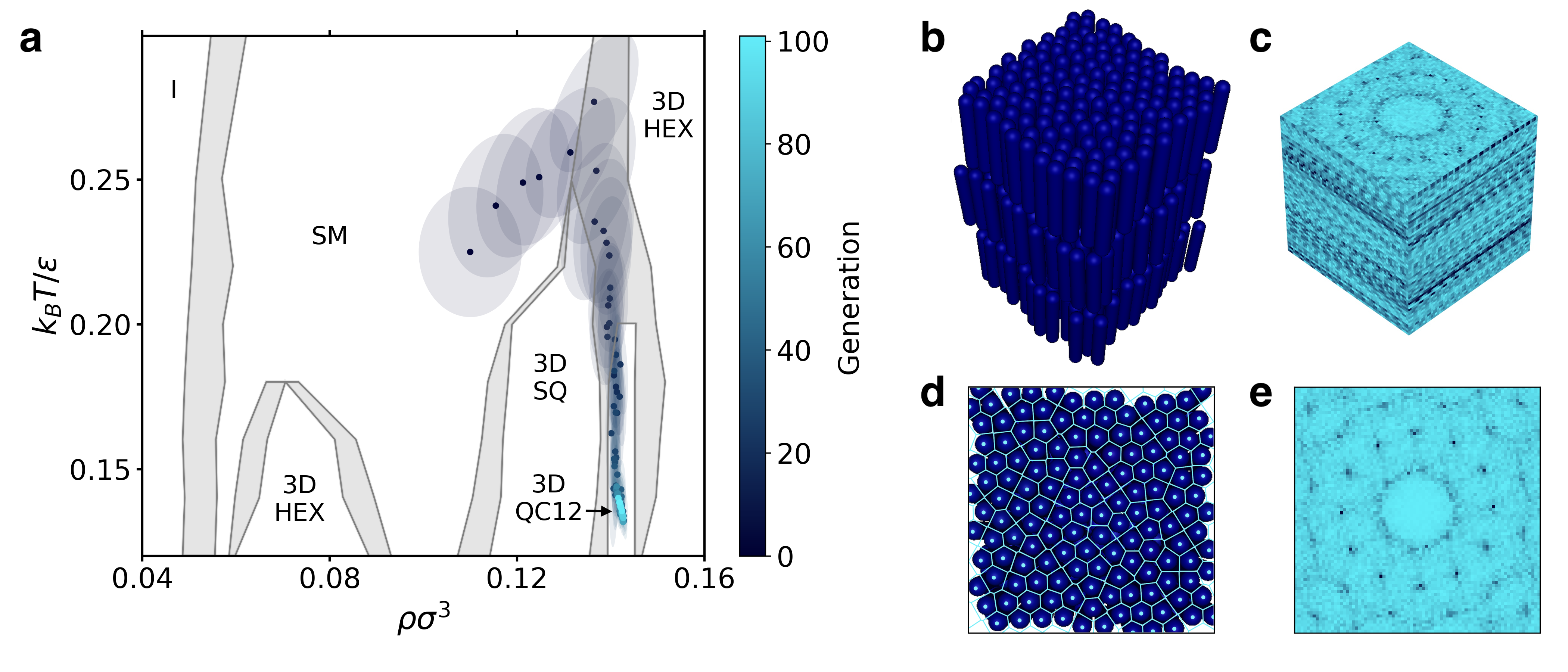}
\caption{{\bf Reverse engineering of the QC12 in a three-dimensional model of soft spherocylinders.} (a) Evolution of the Gaussian distribution in the $\rho\sigma^3-k_BT/\epsilon$ plane. Points and ellipses represent the mean and the covariance matrix (within one standard deviation) of the distribution. The phase diagram in the background is adapted from Ref. \onlinecite{Gerardo}. Coexistence regions are indicated in light gray. (b) Representative snapshot of the 3DQC12 obtained during the last generation, and (c) its three-dimensional diffraction pattern. (d) Top view of the snapshot in (b). The centers of mass and the corresponding Voronoi tessellation are highlighted in a light color. (e) In-layer diffraction pattern of the top view in (d).}
\label{fig:rods}
\end{figure*}

\noindent \textbf{Extension to three-dimensional systems.}
Up to this point, we have shown the efficacy of our method for two-dimensional systems where the scattering pattern is simply a 2D image. Finally, we extend and test our approach on 3D systems. To do so, we  consider a 3D system of rod-like particles, modelled as hard-core spherocylinders with a soft deformable corona. We consider spherocylinders with a length-to-diameter ratio $L/\sigma=5$, interacting via the pair potential in Eq. \ref{eq:scs}, where the center-of-mass distance $r$ is replaced by the minimum distance between two rods $d_m$. Note that $d_m$ depends on both the center-of-mass distance and the relative orientation of the two rods.

The phase behaviour of this system with $k\sigma=10$ and $\delta=1.35\sigma$ has been recently studied in Ref. \cite{Gerardo}. In addition to the standard isotropic (I) and smectic (SM) phases, this model has been shown to stabilize phases consisting of quasi-two-dimensional layers with unconventional symmetries, including  square (3DSQ) and hexagonal (3DHEX) crystals, and a three-dimensional 12-fold quasicrystal (3DQC12). The phase diagram in terms of density and temperature is reported in Fig. \ref{fig:rods}a.

As done in the 2D case, in order to set up our IDM, we train  a CNN to classify all the stable phases of this  system. Note, however, that the inputs of the CNN are now three-dimensional diffraction patterns (see Methods for more details). Again, we find the CNN to be highly effective and able to classify all phases with $100\%$ accuracy.  
The output of the trained CNN is then used to define the fitness for the evolutionary strategy where we target the 3DQC12 phase.

The results of the reverse engineering process are summarized in Fig. \ref{fig:rods}. In particular, Fig. \ref{fig:rods}a shows the evolution of the multivariate Gaussian distribution in the density-temperature plane. Starting with a distribution centered in the SM phase, the mean of the distribution evolves via the  coexistence region of the SM and 3DHEX phase, to the 3DSQ-3DQC12 phase coexistence region, until it converges in the stability region of the 3DQC12 phase. We note that, although the shortest path in parameter space requires the distribution to cross the 3DSQ region, the algorithm actually avoids it, preferring to enter  the coexistence region at high temperature and then move downwards in temperature, where samples with higher fitness are encountered. Surprisingly, this pathway for the formation of QC12 phases was also identified in Ref. \cite{pattabhiraman2017formation}.

A representative snapshot of the 3DQC12 obtained during the last generation along with its 3D diffraction pattern is shown in Figs. \ref{fig:rods}a and \ref{fig:rods}b, respectively. As a further confirmation of the in-layer QC12 arrangement, Figs. \ref{fig:rods}c and \ref{fig:rods}d  show a top view of the same snapshot and the corresponding in-layer 2D diffraction pattern.

The extension of our method to the 3D case is of particular interest from a practical point of view. While a 2D diffraction pattern immediately provides structural information that is easy to read even by eye, the 3D  counterpart is much harder to interpret. For this reason, in order to deal with 3D systems, it is often necessary to project the particles coordinates onto the planes with the relevant symmetries. This aspect becomes irrelevant when using a CNN that naturally processes the full 3D information thanks to its inherent architecture.

\section{Discussion}

Diffraction patterns are used across a multitude of areas in materials science, to understand what structure one is dealing with. In general, this information constitutes a unique signature of each structure, whether it is a crystal, a fluid, a liquid crystal, or a quasicrystal, and shows significant robustness to changes in density and interaction potentials. This can efficiently incorporate all the relevant information of a target phase, and therefore provides a natural order parameter for IDMs.

With the present work we have shown how the use of CNNs as diffraction patterns classifiers, can provide a useful order parameter for the reverse engineering of a multitude of phases. For the above reason, an IDM built on such an order parameter, is not restricted to a specific class of materials, but is instead naturally tailored to reverse engineer multiple colloidal phases, ranging from crystals and quasicrystals, to liquid crystals. 

Our results pave the way to structure optimization and discovery, especially with binary and ternary systems, where the design space becomes even larger due to new system parameters such as size ratio and composition. In these cases, where the present knowledge of phase diagrams and emerging phases is limited, IDMs can prove extremely precious and efficient.

\section{Methods}

\subsection{Convolutional neural networks as a fitness function}
CNNs are a particular type of deep neural networks specifically designed to handle tensorial inputs, such as images. For a detailed description of CNNs see e.g. Ref. \onlinecite{Goodfellow}. In this work, we train a CNN to classify different phases from their diffraction patterns, which are either 2D or 3D images. The output of the CNN is then used to define a fitness function $f$ for the evolutionary strategy. 

More specifically, the CNN takes as input the diffraction pattern of a given configuration, and outputs a vector of real numbers with as many components as the number of phases to distinguish.  Each number in the output is indicative of the probability that the given input corresponds to one of the phases.
We use this CNN to process the configurations saved during each  simulation, and define the fitness of a given sample as
\begin{equation}
f=\bar{P}_{\text{target}}
\end{equation}
where $P_{\text{target}}$ is the probability that the diffraction pattern of a given configuration is classified as the target phase by the CNN, and the bar indicates an average taken over 10 representative configurations saved during the simulation of that sample.

\subsection{Training the convolutional neural networks}
\label{sec:cnntraining}
To train the convolutional neural networks to recognize different phases, we need to perform a number of different steps. Specifically, we first generate a number of real space equilibrium configurations for each phase, and then generate the associated diffraction patterns. In order to reduce  computational time and memory usage, these diffraction patterns are preprocessed before being used to train the convolutional neural networks.  Each of these steps is described in detail in the remainder of this section.

\subsubsection{Generating the training configurations}
The configurations for training the CNNs are generated by performing MC simulations of the 2D HCSS model  \cite{dotera2014mosaic,pattabhiraman2015stability,pattabhiraman2017phase,pattabhiraman2017formation} and the softened-core-shoulder model of spherocylinders in three dimensions (3D) \cite{Gerardo}. 

In 2D, simulations are performed in the isobaric-isothermal ensemble ($NPT$) of a system of $N=256$ particles in a square box of side-length $L$ with periodic boundary conditions. For each of the six phases considered (fluid, HEX, SQ, QC12, QC10, QC18), we run simulations at different state points and collect $10^4$ independent configurations. 

In 3D, simulations are performed in the canonical ensemble ($NVT$) of a system of $N=432$ particles in a rectangular box elongated in the $z$ direction ({\it i.e.}, $L_x=L_y=L$ and $L_z>L$), and with periodic boundary conditions. For each of the five phases considered (I, SM, 3DHEX, 3DSQ, 3DQC12), we run simulations at different state points and collect $5\cdot 10^3$ independent configurations. 

\subsubsection{Generating the diffraction patterns}
Diffraction patterns for each configuration are evaluated using 
\begin{equation}
S(\mathbf{k}) = \frac{1}{N} \rho(\mathbf{k})\rho(\mathbf{-k}),
\end{equation}
where $\rho(\mathbf{k})=\sum_{j=1}^{N} e^{-i\mathbf{k}\cdot\mathbf{r}_j}$ is the Fourier transform of the density, $\mathbf{r}_j$ is the position of particle $j$, and $\mathbf{k}$ is a wave vector. In 2D, the $\mathbf{k}$ vectors are chosen by $\mathbf{k}=\frac{2\pi}{L}(n_x, n_y)$, where $n_x$ and $n_y$ are two integers in the interval $[-64,64]$. As a result, the 2D diffraction patterns considered in this work are built on a $129\times 129$ grid.
In 3D, the $\mathbf{k}$ vectors are chosen by $\mathbf{k}=2\pi(\frac{n_x}{L}, \frac{n_y}{L}, \frac{n_z}{L_z})$, with $n_x,n_y,n_x \in [-32,32]$, resulting in a $65\times 65 \times 65$ grid.

\begin{figure}[!h]
\center
\includegraphics[width=\linewidth]{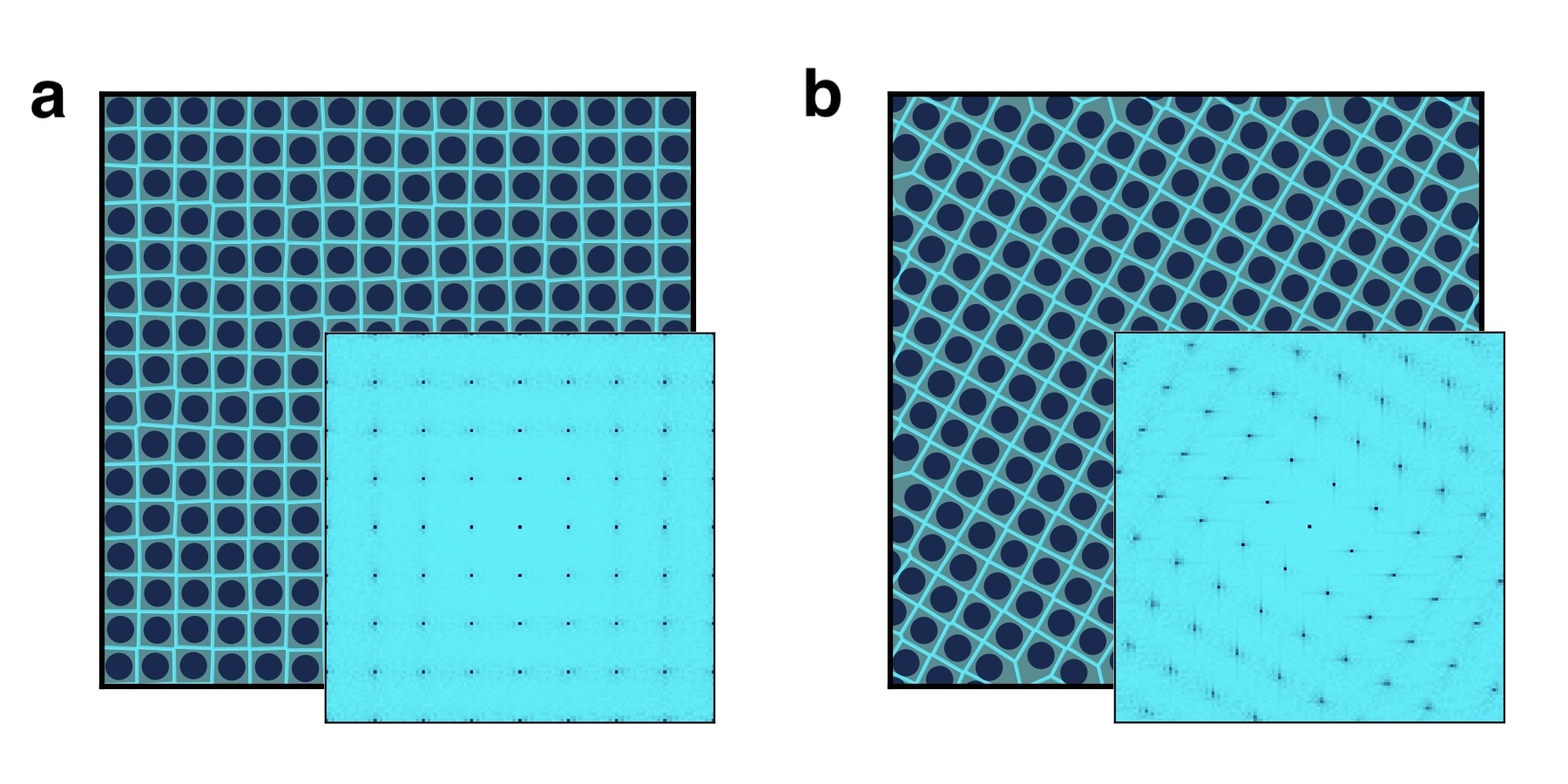}
\caption{{\bf Data transformation.} (a) Snapshot and diffraction pattern of a square crystal in its original orientation. (b) Same snapshot and diffraction pattern as (a) after a rotation by a $\pi/6$ angle. Note that the rotation is performed in real space.} 
\label{fig:rotation}
\end{figure}

While diffraction patterns are by definition translationally invariant, they are not invariant to rotations. However, we must ensure that the CNNs are able to classify the desired phases regardless of their orientation. To this end, each training configuration is rotated by a random angle before evaluating its diffraction pattern. A representation of this transformation in the 2D case is shown in Fig. \ref{fig:rotation}. In the 3D case, given the inherent symmetry of the model of spherocylinders considered, we randomly rotate each configuration around the $z$ axis (which always corresponds to the elongated axis of the box). In a more general case, one could perform random rotations around a randomly selected axis. Note that, to rotate a configuration, we first create a larger copy of the system by copying the original simulation box in all directions. We then rotate this larger copy of the system, and finally take a portion of it of the same size as the original simulation box.

The sets of diffraction patterns obtained after having rotated each configuration are finally used to build the data sets for training the CNNs.

\subsubsection{Preprocessing}
\begin{figure}[!h]
\center
\includegraphics[width=\linewidth]{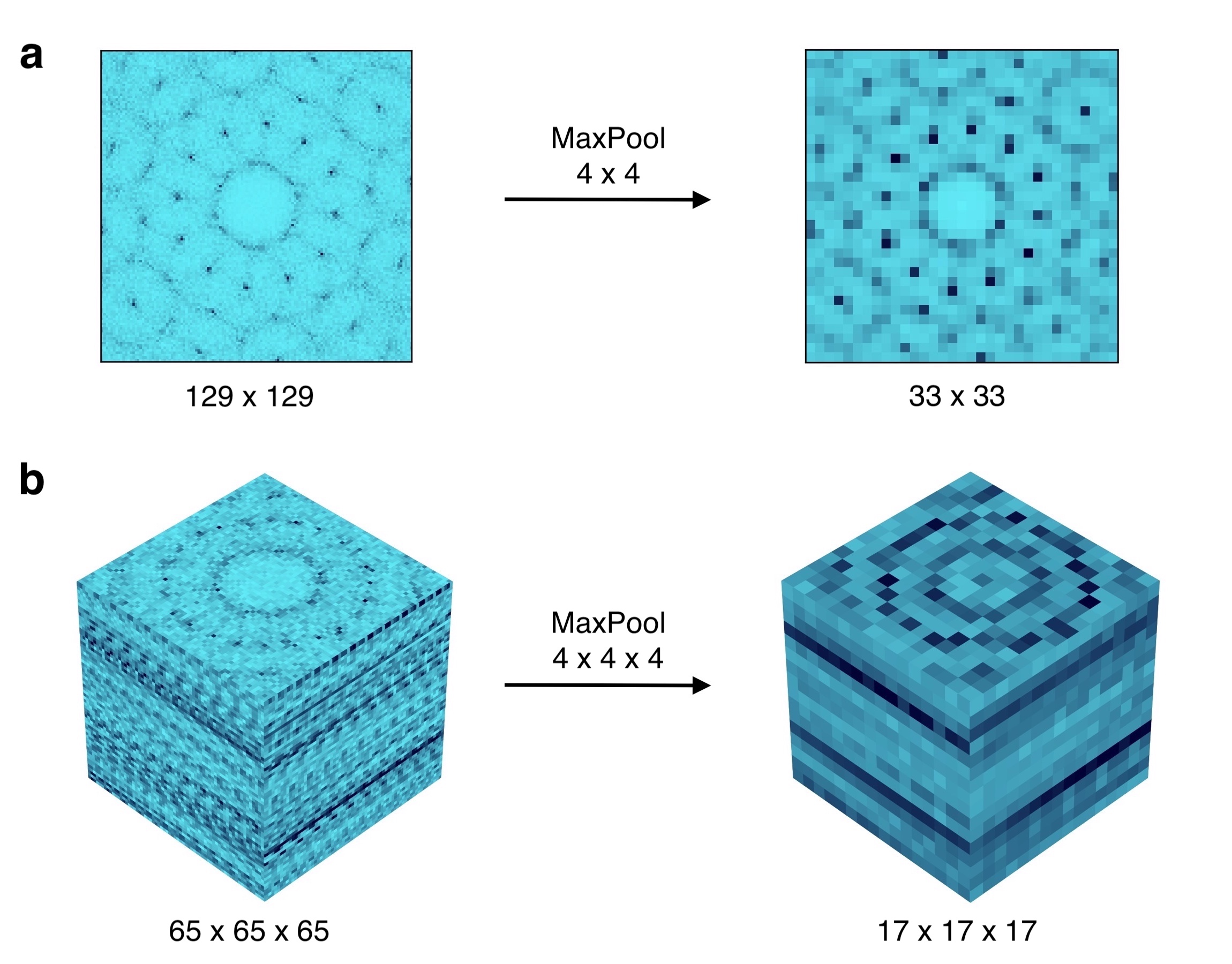}
\caption{{\bf Preprocessing.} The size of the diffraction pattern of a QC12 in (a) 2D and (b) 3D is reduced through a MaxPooling filter.} 
\label{fig:preprocessing}
\end{figure}

In order to increase the overall efficiency, the diffraction patterns undergo a final preprocessing step before being used as the input of the CNNs. In particular, each diffraction pattern passes through a MaxPooling filter, that effectively reduces the input size by a factor 4 in each dimension. The effect of this transformation is shown in Fig. \ref{fig:preprocessing} for both the (a) 2D, and (b) 3D cases. 
Note that this is not a necessary step of the algorithm and its only purpose is to increase the efficiency of the method in terms of computational time and memory usage. With such a preprocessing, the CNNs used here can be trained within one hour on the CPU of a modern laptop. 

\subsubsection{Neural network architecture}
The CNNs used in this work are composed of two convolutional layers for feature extraction, and a fully-connected part with one hidden layer for the final classification.
\begin{figure*}
\center
\includegraphics[width=\linewidth]{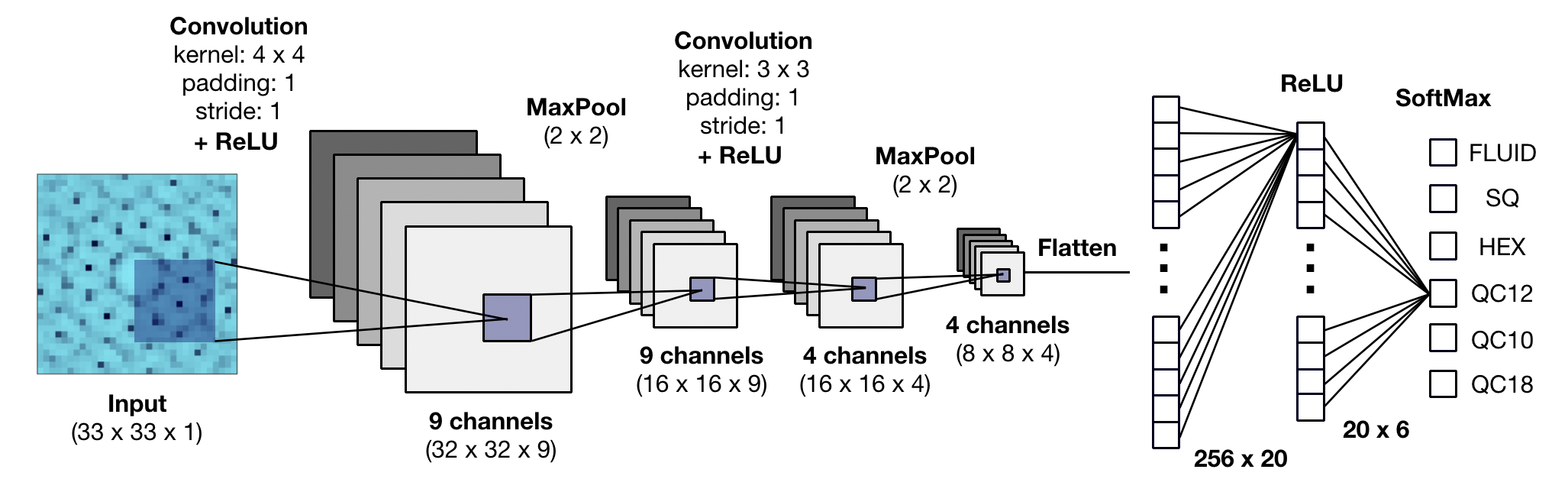}
\caption{{\bf Representation of the 2D convolutional neural network.} The network is composed of two convolutional layers for feature extraction, and a fully-connected part with one hidden layer for the final classification. All details about kernels, layer size, and activation functions are also shown.} 
\label{fig:cnn}
\end{figure*}
The architecture of the 2D CNN is shown in Fig. \ref{fig:cnn}. As shown in the figure, each convolutional layer performs three operations on the input: a convolution, a non-linear transformation through a ReLU activation function, and a downsampling operation through a $2\times 2$ MaxPooling layer. In the following, we give all the details about the network parameters.

The first convolutional layer has one input channel (i.e., the diffraction pattern to process) and nine output channels (i.e., the extracted features). As indicated in Fig. \ref{fig:cnn}, the kernels used in this layer have a size $s=4\times 4$, padding $p=1$, and stride $s=1$.
The second convolutional layer has nine input channels and four output channels, and the kernels of this layer have a size $s=3\times3$, padding $p=1$, and stride $s=1$. 
The output of the second convolutional layer is stacked and flattened, in order to be used as the input of the fully-connected part of the network. The latter consists of a hidden layer of dimension $20$ with a ReLU activation function, and an output layer with a SoftMax activation function. The size of the output layer is equal to the number of phases we wish to distinguish, which is 6 in the 2D case.

The 3D CNN has almost the same structure as the 2D one, with the only exception being that the convolutional kernels are extended to three dimensions (e.g. a $3\times 3$ kernel in 2D becomes a $3\times 3\times 3$ kernel in 3D), and the output layer has a dimension of $5$ (we consider $5$ phases in the 3D system).

\subsubsection{Training}
The parameters of the CNNs are optimized by minimizing the cross-entropy loss with the addition of a weight decay regularization term \cite{Bishop,Bishop2}. Specifically, the loss is minimized with the Adam optimizer \cite{kingma2014adam}, a learning rate of $10^{-4}$, and a PyTorch implementation \cite{paszke2019pytorch}. Early stopping is also applied in order to prevent overfitting.

\subsection{Workflow of the CMA evolutionary strategy}
The CMA evolutionary strategy optimizes iteratively the design parameters across successive generations.   At each generation, we draw $n$ samples  from a multivariate Gaussian distribution, whose dimension $D$ corresponds to the number of parameters we wish to optimize. Subsequently, we evaluate the fitness function $f$ of the generated samples, we order the samples in ascending order based on their fitness, and we pick the set $\mathbf{X}$ of the best $k$ samples.
Finally, the mean $\vec \mu $ (a $D$-dimensional vector) and the covariance matrix $\mathbf{\Sigma} = \sigma^2\mathbf{C}$ of the Gaussian distribution are updated using the following equations:
\begin{equation}
\begin{aligned}
{\mu_i}^\prime &= {\mu_i} + \sum_{x\in\mathbf{X}}w(x)(\lambda_i (x) - \mu_i) \\
{q_i}^\prime &= (1-c_1)q_i + c_2{\sqrt{\Sigma^{-1}}}_{ij}({\mu_j}^\prime - \mu_j)\\
{p_i}^\prime &= (1-c_3)p_i + c_4({\mu_i}^\prime - \mu_i)\\
{C_{ij}}^\prime &= (1-c_5-c_6)C_{ij} + c_6{p_i}^\prime{p_j}^\prime \\ 
 \quad &\hspace{-0.5cm}+ c_5\sum_{x\in\mathbf{X}}w(x)\left(\frac{\lambda_i (x) - \mu_i}{\sigma}\frac{\lambda_j (x) - \mu_j}{\sigma} - C_{ij} \right) \\
\sigma^\prime &= \sigma \mathrm{\, exp\,}[c_7(\frac{\parallel{\vec{q^\prime}} \parallel}{\langle \parallel N(0,I)\parallel \rangle} - 1)]
\end{aligned}
\end{equation}
where $\mathbf{X}$ denotes the set of the $k$ best samples consisting of multiple configurations obtained for $k$ different parameter sets (denoted by $\lambda_i(x)$),  $w(x)$ is the normalized distribution of weights based on the fitness of the samples, and $c_i$'s are free parameters. We choose $w(x) \propto \log(k+1) - \log(m)$, where $m$ is the rank index of sample $x$ ($m = 1$ for the configuration with the largest $f$ value). 
$\vec{q}$ and $\vec{p}$ are additional $D$-dimensional vectors that control, respectively, the changes in amplitude and directionality of the covariance matrix. Additionally, $ \langle \parallel N(0,I)\parallel \rangle $ is the average length of a vector drawn from a multivariate Gaussian distribution centered in the origin and where the covariance matrix is the identity matrix. In the present work, we use $n = 10$ and $k = 5$ for all cases where we optimize two parameters, {\it i.e.} $D=2$. When optimizing three parameters ($D=3$), we use instead $n = 20$ and $k = 8$ in order to guarantee a faster exploration of the phase space. For the first generation, we initialize $\vec q $ and $\vec p $ as null vectors. Moreover, since we do not assume any \emph{a priori} correlation between the different tuning parameters, the initial form of the covariance matrix $\mathbf{\Sigma}$ is diagonal. Finally, all the free parameters $c_i$ of the CMA-ES are set equal to 0.2, as proposed in Ref. \onlinecite{hansen2006towards}.

\subsection{Simulation Details}
At every generation, we perform  Monte Carlo simulations for each of the sets of parameters drawn from the multivariate Gaussian distribution. In each simulation, after the system has equilibrated, we save $10$ independent configurations, which are then used to evaluate the fitness of the samples.

For the HCSS model, simulations are performed in the isobaric-isothermal ensemble in a two-dimensional box with periodic boundary conditions, and with a system size of $N=256$ particles. In all cases, the system is initialized in a disordered, low-density, configuration.

For the SCS model, simulations are performed both in the canonical and the  isobaric-isothermal ensembles in a two-dimensional box with periodic boundary conditions, and with a system size of $N=256$ particles. In all cases, the system is initialized in a random configuration.

For the three-dimensional system of spherocylinders, simulations are performed in the canonical ensemble
considering a system size of $N=432$ particles in a three-dimensional rectangular box elongated in the $z$ direction. In this case, all simulations are initialized in a smectic configuration.

\section{Data availability}

The data associated with this research is available upon request.

\section{Code availability}

The simulation and analysis codes associated with this research are available upon request.

\bibliography{ref}

\begin{thebibliography}{10}
\expandafter\ifx\csname url\endcsname\relax
  \def\url#1{\texttt{#1}}\fi
\expandafter\ifx\csname urlprefix\endcsname\relax\def\urlprefix{URL }\fi
\providecommand{\bibinfo}[2]{#2}
\providecommand{\eprint}[2][]{\url{#2}}

\bibitem{glotzer2007anisotropy}
\bibinfo{author}{Glotzer, S.~C.} \& \bibinfo{author}{Solomon, M.~J.}
\newblock \bibinfo{title}{Anisotropy of building blocks and their assembly into
  complex structures}.
\newblock \emph{\bibinfo{journal}{Nat. Mater}} \textbf{\bibinfo{volume}{6}},
  \bibinfo{pages}{557--562} (\bibinfo{year}{2007}).

\bibitem{sacanna2011shape}
\bibinfo{author}{Sacanna, S.} \& \bibinfo{author}{Pine, D.~J.}
\newblock \bibinfo{title}{Shape-anisotropic colloids: Building blocks for
  complex assemblies}.
\newblock \emph{\bibinfo{journal}{Curr. Opin. Colloid Interface Sci.}}
  \textbf{\bibinfo{volume}{16}}, \bibinfo{pages}{96--105}
  (\bibinfo{year}{2011}).

\bibitem{miszta2011hierarchical}
\bibinfo{author}{Miszta, K.} \emph{et~al.}
\newblock \bibinfo{title}{Hierarchical self-assembly of suspended branched
  colloidal nanocrystals into superlattice structures}.
\newblock \emph{\bibinfo{journal}{Nat. Mater}} \textbf{\bibinfo{volume}{10}},
  \bibinfo{pages}{872--876} (\bibinfo{year}{2011}).

\bibitem{boles2016self}
\bibinfo{author}{Boles, M.~A.}, \bibinfo{author}{Engel, M.} \&
  \bibinfo{author}{Talapin, D.~V.}
\newblock \bibinfo{title}{Self-assembly of colloidal nanocrystals: From
  intricate structures to functional materials}.
\newblock \emph{\bibinfo{journal}{Chem. Rev.}} \textbf{\bibinfo{volume}{116}},
  \bibinfo{pages}{11220--11289} (\bibinfo{year}{2016}).

\bibitem{he2020colloidal}
\bibinfo{author}{He, M.} \emph{et~al.}
\newblock \bibinfo{title}{Colloidal diamond}.
\newblock \emph{\bibinfo{journal}{Nature}} \textbf{\bibinfo{volume}{585}},
  \bibinfo{pages}{524--529} (\bibinfo{year}{2020}).

\bibitem{rechtsman2005optimized}
\bibinfo{author}{Rechtsman, M.~C.}, \bibinfo{author}{Stillinger, F.~H.} \&
  \bibinfo{author}{Torquato, S.}
\newblock \bibinfo{title}{Optimized interactions for targeted self-assembly:
  application to a honeycomb lattice}.
\newblock \emph{\bibinfo{journal}{Phys. Rev. Lett.}}
  \textbf{\bibinfo{volume}{95}}, \bibinfo{pages}{228301}
  (\bibinfo{year}{2005}).

\bibitem{florescu2009designer}
\bibinfo{author}{Florescu, M.}, \bibinfo{author}{Torquato, S.} \&
  \bibinfo{author}{Steinhardt, P.~J.}
\newblock \bibinfo{title}{Designer disordered materials with large, complete
  photonic band gaps}.
\newblock \emph{\bibinfo{journal}{Proc. Natl. Acad. Sci. U.S.A}}
  \textbf{\bibinfo{volume}{106}}, \bibinfo{pages}{20658--20663}
  (\bibinfo{year}{2009}).

\bibitem{miskin2013adapting}
\bibinfo{author}{Miskin, M.~Z.} \& \bibinfo{author}{Jaeger, H.~M.}
\newblock \bibinfo{title}{Adapting granular materials through artificial
  evolution}.
\newblock \emph{\bibinfo{journal}{Nat. Mater}} \textbf{\bibinfo{volume}{12}},
  \bibinfo{pages}{326--331} (\bibinfo{year}{2013}).

\bibitem{jain2014inverse}
\bibinfo{author}{Jain, A.}, \bibinfo{author}{Bollinger, J.~A.} \&
  \bibinfo{author}{Truskett, T.~M.}
\newblock \bibinfo{title}{Inverse methods for material design}
  (\bibinfo{year}{2014}).

\bibitem{long2018rational}
\bibinfo{author}{Long, A.~W.} \& \bibinfo{author}{Ferguson, A.~L.}
\newblock \bibinfo{title}{Rational design of patchy colloids via landscape
  engineering}.
\newblock \emph{\bibinfo{journal}{Mol. Syst. Des. Eng.}}
  \textbf{\bibinfo{volume}{3}}, \bibinfo{pages}{49--65} (\bibinfo{year}{2018}).

\bibitem{kumar2019inverse}
\bibinfo{author}{Kumar, R.}, \bibinfo{author}{Coli, G.~M.},
  \bibinfo{author}{Dijkstra, M.} \& \bibinfo{author}{Sastry, S.}
\newblock \bibinfo{title}{Inverse design of charged colloidal particle
  interactions for self assembly into specified crystal structures}.
\newblock \emph{\bibinfo{journal}{J. Chem. Phys.}}
  \textbf{\bibinfo{volume}{151}}, \bibinfo{pages}{084109}
  (\bibinfo{year}{2019}).

\bibitem{khadilkar2017inverse}
\bibinfo{author}{Khadilkar, M.~R.}, \bibinfo{author}{Paradiso, S.},
  \bibinfo{author}{Delaney, K.~T.} \& \bibinfo{author}{Fredrickson, G.~H.}
\newblock \bibinfo{title}{Inverse design of bulk morphologies in multiblock
  polymers using particle swarm optimization}.
\newblock \emph{\bibinfo{journal}{Macromolecules}}
  \textbf{\bibinfo{volume}{50}}, \bibinfo{pages}{6702--6709}
  (\bibinfo{year}{2017}).

\bibitem{miskin2016turning}
\bibinfo{author}{Miskin, M.~Z.}, \bibinfo{author}{Khaira, G.},
  \bibinfo{author}{de~Pablo, J.~J.} \& \bibinfo{author}{Jaeger, H.~M.}
\newblock \bibinfo{title}{Turning statistical physics models into materials
  design engines}.
\newblock \emph{\bibinfo{journal}{Proc. Natl. Acad. Sci. U.S.A}}
  \textbf{\bibinfo{volume}{113}}, \bibinfo{pages}{34--39}
  (\bibinfo{year}{2016}).

\bibitem{geng2019engineering}
\bibinfo{author}{Geng, Y.}, \bibinfo{author}{van Anders, G.},
  \bibinfo{author}{Dodd, P.~M.}, \bibinfo{author}{Dshemuchadse, J.} \&
  \bibinfo{author}{Glotzer, S.~C.}
\newblock \bibinfo{title}{Engineering entropy for the inverse design of
  colloidal crystals from hard shapes}.
\newblock \emph{\bibinfo{journal}{Sci. Adv.}} \textbf{\bibinfo{volume}{5}}
  (\bibinfo{year}{2019}).

\bibitem{hansen2006towards}
\bibinfo{author}{Hansen, N.}
\newblock \bibinfo{title}{Towards a new evolutionary computation}.
\newblock \emph{\bibinfo{journal}{Stud. Fuzziness Soft Comput.}}
  \textbf{\bibinfo{volume}{192}}, \bibinfo{pages}{75--102}
  (\bibinfo{year}{2006}).

\bibitem{jain2013inverse}
\bibinfo{author}{Jain, A.}, \bibinfo{author}{Errington, J.~R.} \&
  \bibinfo{author}{Truskett, T.~M.}
\newblock \bibinfo{title}{Inverse design of simple pairwise interactions with
  low-coordinated 3d lattice ground states}.
\newblock \emph{\bibinfo{journal}{Soft Matter}} \textbf{\bibinfo{volume}{9}},
  \bibinfo{pages}{3866--3870} (\bibinfo{year}{2013}).

\bibitem{van2015digital}
\bibinfo{author}{van Anders, G.}, \bibinfo{author}{Klotsa, D.},
  \bibinfo{author}{Karas, A.~S.}, \bibinfo{author}{Dodd, P.~M.} \&
  \bibinfo{author}{Glotzer, S.~C.}
\newblock \bibinfo{title}{Digital alchemy for materials design: Colloids and
  beyond}.
\newblock \emph{\bibinfo{journal}{ACS nano}} \textbf{\bibinfo{volume}{9}},
  \bibinfo{pages}{9542--9553} (\bibinfo{year}{2015}).

\bibitem{jadrich2015equilibrium}
\bibinfo{author}{Jadrich, R.~B.}, \bibinfo{author}{Bollinger, J.~A.},
  \bibinfo{author}{Lindquist, B.~A.} \& \bibinfo{author}{Truskett, T.~M.}
\newblock \bibinfo{title}{Equilibrium cluster fluids: Pair interactions via
  inverse design}.
\newblock \emph{\bibinfo{journal}{Soft Matter}} \textbf{\bibinfo{volume}{11}},
  \bibinfo{pages}{9342--9354} (\bibinfo{year}{2015}).

\bibitem{pineros2018inverse}
\bibinfo{author}{Pi{\~n}eros, W.~D.}, \bibinfo{author}{Lindquist, B.~A.},
  \bibinfo{author}{Jadrich, R.~B.} \& \bibinfo{author}{Truskett, T.~M.}
\newblock \bibinfo{title}{Inverse design of multicomponent assemblies}.
\newblock \emph{\bibinfo{journal}{J. Chem. Phys.}}
  \textbf{\bibinfo{volume}{148}}, \bibinfo{pages}{104509}
  (\bibinfo{year}{2018}).

\bibitem{sherman2020inverse}
\bibinfo{author}{Sherman, Z.~M.}, \bibinfo{author}{Howard, M.~P.},
  \bibinfo{author}{Lindquist, B.~A.}, \bibinfo{author}{Jadrich, R.~B.} \&
  \bibinfo{author}{Truskett, T.~M.}
\newblock \bibinfo{title}{Inverse methods for design of soft materials}.
\newblock \emph{\bibinfo{journal}{J. Chem. Phys.}}
  \textbf{\bibinfo{volume}{152}}, \bibinfo{pages}{140902}
  (\bibinfo{year}{2020}).

\bibitem{adorf2018inverse}
\bibinfo{author}{Adorf, C.~S.}, \bibinfo{author}{Antonaglia, J.},
  \bibinfo{author}{Dshemuchadse, J.} \& \bibinfo{author}{Glotzer, S.~C.}
\newblock \bibinfo{title}{Inverse design of simple pair potentials for the
  self-assembly of complex structures}.
\newblock \emph{\bibinfo{journal}{J. Chem. Phys.}}
  \textbf{\bibinfo{volume}{149}}, \bibinfo{pages}{204102}
  (\bibinfo{year}{2018}).

\bibitem{barkan2011stability}
\bibinfo{author}{Barkan, K.}, \bibinfo{author}{Diamant, H.} \&
  \bibinfo{author}{Lifshitz, R.}
\newblock \bibinfo{title}{Stability of quasicrystals composed of soft isotropic
  particles}.
\newblock \emph{\bibinfo{journal}{Phys. Rev. B}} \textbf{\bibinfo{volume}{83}},
  \bibinfo{pages}{172201} (\bibinfo{year}{2011}).

\bibitem{dotera2014mosaic}
\bibinfo{author}{Dotera, T.}, \bibinfo{author}{Oshiro, T.} \&
  \bibinfo{author}{Ziherl, P.}
\newblock \bibinfo{title}{Mosaic two-lengthscale quasicrystals}.
\newblock \emph{\bibinfo{journal}{Nature}} \textbf{\bibinfo{volume}{506}},
  \bibinfo{pages}{208--211} (\bibinfo{year}{2014}).

\bibitem{pattabhiraman2015stability}
\bibinfo{author}{Pattabhiraman, H.}, \bibinfo{author}{Gantapara, A.~P.} \&
  \bibinfo{author}{Dijkstra, M.}
\newblock \bibinfo{title}{On the stability of a quasicrystal and its
  crystalline approximant in a system of hard disks with a soft corona}.
\newblock \emph{\bibinfo{journal}{J. Chem. Phys.}}
  \textbf{\bibinfo{volume}{143}}, \bibinfo{pages}{164905}
  (\bibinfo{year}{2015}).

\bibitem{pattabhiraman2017phase}
\bibinfo{author}{Pattabhiraman, H.} \& \bibinfo{author}{Dijkstra, M.}
\newblock \bibinfo{title}{Phase behaviour of quasicrystal forming systems of
  core-corona particles}.
\newblock \emph{\bibinfo{journal}{J. Chem. Phys.}}
  \textbf{\bibinfo{volume}{146}}, \bibinfo{pages}{114901}
  (\bibinfo{year}{2017}).

\bibitem{pattabhiraman2017formation}
\bibinfo{author}{Pattabhiraman, H.} \& \bibinfo{author}{Dijkstra, M.}
\newblock \bibinfo{title}{On the formation of stripe, sigma, and honeycomb
  phases in a core--corona system}.
\newblock \emph{\bibinfo{journal}{Soft Matter}} \textbf{\bibinfo{volume}{13}},
  \bibinfo{pages}{4418--4432} (\bibinfo{year}{2017}).

\bibitem{padilla2020phase}
\bibinfo{author}{Padilla, L.~A.} \&
  \bibinfo{author}{Ram{\'\i}rez-Hern{\'a}ndez, A.}
\newblock \bibinfo{title}{Phase behavior of a two-dimensional core-softened
  system: new physical insights}.
\newblock \emph{\bibinfo{journal}{J. Phys. Condens. Matter}}
  \textbf{\bibinfo{volume}{32}}, \bibinfo{pages}{275103}
  (\bibinfo{year}{2020}).

\bibitem{kryuchkov2018complex}
\bibinfo{author}{Kryuchkov, N.~P.}, \bibinfo{author}{Yurchenko, S.~O.},
  \bibinfo{author}{Fomin, Y.~D.}, \bibinfo{author}{Tsiok, E.~N.} \&
  \bibinfo{author}{Ryzhov, V.~N.}
\newblock \bibinfo{title}{Complex crystalline structures in a two-dimensional
  core-softened system}.
\newblock \emph{\bibinfo{journal}{Soft Matter}} \textbf{\bibinfo{volume}{14}},
  \bibinfo{pages}{2152--2162} (\bibinfo{year}{2018}).

\bibitem{Gerardo}
\bibinfo{author}{Campos-Villalobos, G.}, \bibinfo{author}{Dijkstra, M.} \&
  \bibinfo{author}{Patti, A.}
\newblock \bibinfo{title}{Nonconventional phases of colloidal nanorods with a
  soft corona}.
\newblock \emph{\bibinfo{journal}{Phys. Rev. Lett.}}
  \textbf{\bibinfo{volume}{126}}, \bibinfo{pages}{158001}
  (\bibinfo{year}{2021}).

\bibitem{Goodfellow}
\bibinfo{author}{Goodfellow, I.}, \bibinfo{author}{Bengio, Y.} \&
  \bibinfo{author}{Courville, A.}
\newblock \emph{\bibinfo{title}{Deep Learning}} (\bibinfo{publisher}{The MIT
  Press}, \bibinfo{year}{2016}).

\bibitem{Bishop}
\bibinfo{author}{Bishop, C.~M.}
\newblock \emph{\bibinfo{title}{Neural Networks for Pattern Recognition}}
  (\bibinfo{publisher}{Oxford University Press, Inc.}, \bibinfo{address}{New
  York, NY, USA}, \bibinfo{year}{1995}).

\bibitem{Bishop2}
\bibinfo{author}{Bishop, C.~M.}
\newblock \emph{\bibinfo{title}{Pattern recognition and machine learning}}
  (\bibinfo{publisher}{Springer}, \bibinfo{year}{2006}).

\bibitem{kingma2014adam}
\bibinfo{author}{Kingma, D.~P.} \& \bibinfo{author}{Ba, J.}
\newblock \bibinfo{title}{Adam: A method for stochastic optimization}.
\newblock \emph{\bibinfo{journal}{arXiv preprint arXiv:1412.6980}}
  (\bibinfo{year}{2014}).

\bibitem{paszke2019pytorch}
\bibinfo{author}{Paszke, A.} \emph{et~al.}
\newblock \bibinfo{title}{Pytorch: An imperative style, high-performance deep
  learning library}.
\newblock In \emph{\bibinfo{booktitle}{Advances in neural information
  processing systems}}, \bibinfo{pages}{8026--8037} (\bibinfo{year}{2019}).

\end{thebibliography}


\begin{thebibliography}{1}
\expandafter\ifx\csname url\endcsname\relax
  \def\url#1{\texttt{#1}}\fi
\expandafter\ifx\csname urlprefix\endcsname\relax\def\urlprefix{URL }\fi
\providecommand{\bibinfo}[2]{#2}
\providecommand{\eprint}[2][]{\url{#2}}

\bibitem{pattabhiraman2015stability}
\bibinfo{author}{Pattabhiraman, H.}, \bibinfo{author}{Gantapara, A.~P.} \&
  \bibinfo{author}{Dijkstra, M.}
\newblock \bibinfo{title}{On the stability of a quasicrystal and its
  crystalline approximant in a system of hard disks with a soft corona}.
\newblock \emph{\bibinfo{journal}{J. Chem. Phys.}}
  \textbf{\bibinfo{volume}{143}}, \bibinfo{pages}{164905}
  (\bibinfo{year}{2015}).

\end{thebibliography}

\section{Acknowledgements}
All authors acknowledge financial support from NWO (Grants no. 16DDS003 and 16DDS004).

\section{Author Contributions}
M.D. initiated the inverse design project of quasicrystals, liquid crystals, and crystals, and supervised G.M.C. L.F. supervised E.B. G.M.C. and E.B. performed the research and contributed equally to this work.  All authors co-wrote the manuscript, and discussed the text and interpretation of the results.
\end{document}



\makeatletter
\def\fnum@figure{\figurename\nobreakspace\textbf{\thefigure}}
\makeatother

\renewcommand{\figurename}{\textbf{Supplementary Figure}}
\renewcommand{\thesection}{}

 \renewcommand{\bibsection}{\section{References}}

\titleformat{\section}[display]
  {\centering\normalfont\scshape \bfseries}{
  \MakeUppercase{#1}}{0em}{}
  
\titlespacing\section{0pt}{20pt plus 4pt minus 2pt}{0pt plus 2pt minus 2pt}

\title{Supplementary Information\\
Inverse design of soft materials via a deep-learning-based evolutionary strategy}


\author{Gabriele M. Coli}
\affiliation{Soft Condensed Matter, Debye Institute of Nanomaterials Science, Utrecht University, Utrecht, Netherlands}

\author{Emanuele Boattini}
\affiliation{Soft Condensed Matter, Debye Institute of Nanomaterials Science, Utrecht University, Utrecht, Netherlands}

\author{Laura Filion}
\affiliation{Soft Condensed Matter, Debye Institute of Nanomaterials Science, Utrecht University, Utrecht, Netherlands}

\author{Marjolein Dijkstra}
\affiliation{Soft Condensed Matter, Debye Institute of Nanomaterials Science, Utrecht University, Utrecht, Netherlands}

\maketitle




\onecolumngrid
\vspace{2cm} 
\begin{center}
{\bf \MakeUppercase{Contents}}
\end{center}
\vspace{-1cm}
\tableofcontents
\vspace{\columnsep}

\null
\clearpage

\section{Supplementary note 1: Reverse engineering of the QC12 starting from different initial conditions}
In the main text, we have shown a trajectory of the reverse engineering of the QC12 in the HCSS model, starting from a specific state point in the region of stability of the fluid phase. Here, we explore whether the performance of the method is affected by a different choice of the initial conditions. To this end, we perform additional trajectories of the reverse engineering of the QC12,  starting with a Gaussian distribution centered at different state points, i.e. in the fluid phase, the SQ phase, the HEX phase at relatively high temperature and low pressure, and the HEX phase at relatively low temperature and high pressure.

\begin{figure*}[!h]
\includegraphics[width=1.0\linewidth]{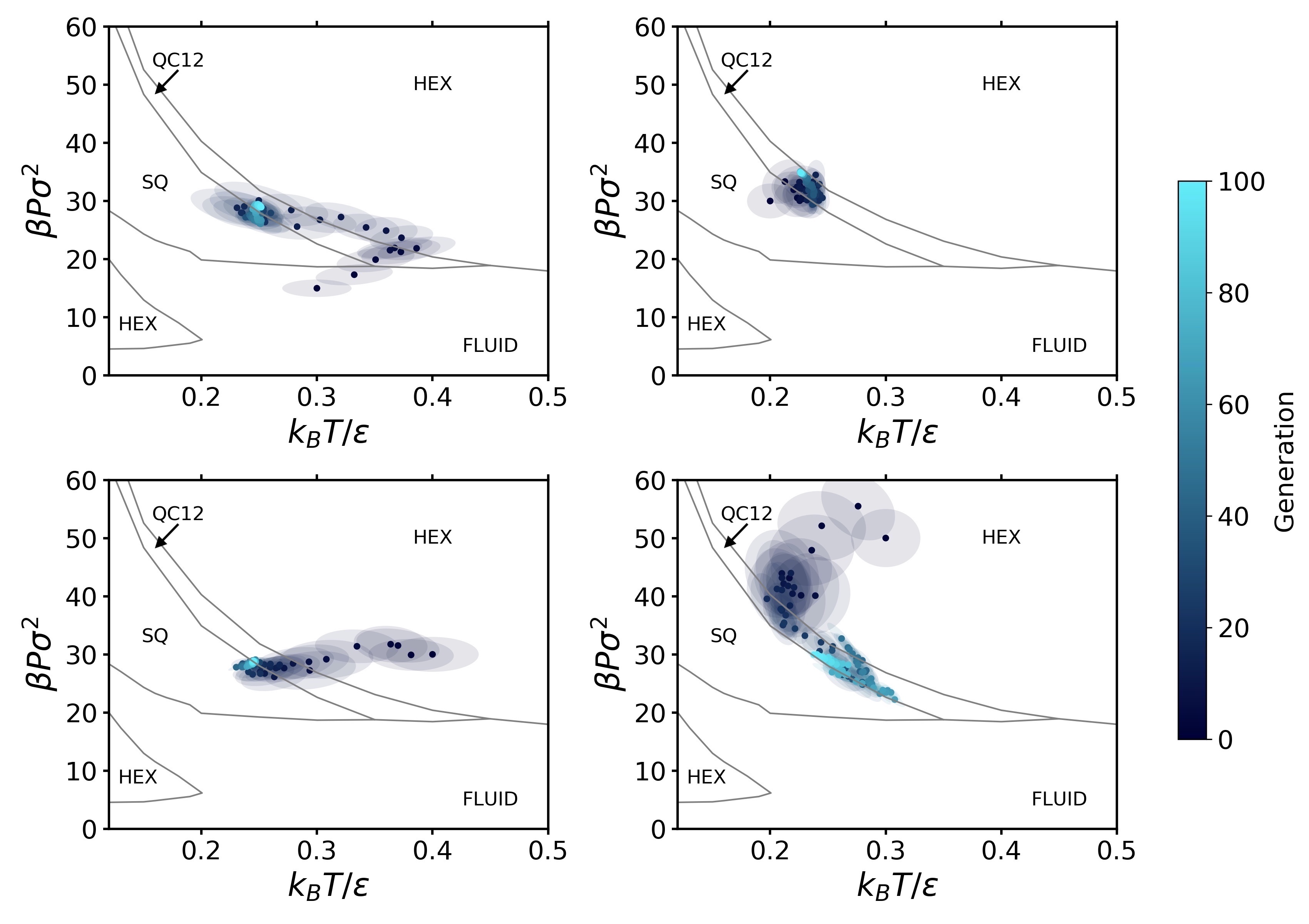}
\caption{{\bf Reverse engineering of the QC12 starting from different initial conditions}. Four different trajectories showing the evolution of the Gaussian distribution in the $k_BT/\epsilon-\beta P\sigma^2$ plane. Each trajectory, is initialized with the Gaussian centered at different state points, i.e. in the fluid phase (left top), the SQ phase (right top), the HEX phase at relatively high temperature and low pressure (left bottom), and the HEX phase at relatively low temperature and high pressure (right bottom). Points and ellipses represent the mean and the covariance matrix (within one standard deviation) of the distribution. The phase diagram in the background is adapted from Ref. \onlinecite{pattabhiraman2015stability}.} 
\label{fig:additionalHCSS}
\end{figure*}

Supplementary Figure \ref{fig:additionalHCSS} shows the results of the reverse engineering process obtained from four, distinct, initial state points. In all cases, the mean of the parameters distribution converges within the region of stability of the target QC12, clearly showing that the performance is not affected by the particular choice made for the initial conditions.

\clearpage

\section{Supplementary note 2: Reverse engineering of the hexagonal crystal in the HCSS model}

In the main text, we focused on reverse engineering  QCs. However, in principle, the exact same method can be used to reverse engineer any phase that was included in the data set for training the CNN, simply by changing the definition of the fitness. As an example, in Supplementary Figure \ref{fig:hex} we report the results of the reverse engineering of the HEX phase in the HCSS model.

\begin{figure*}[!h]
\includegraphics[width=1.0\linewidth]{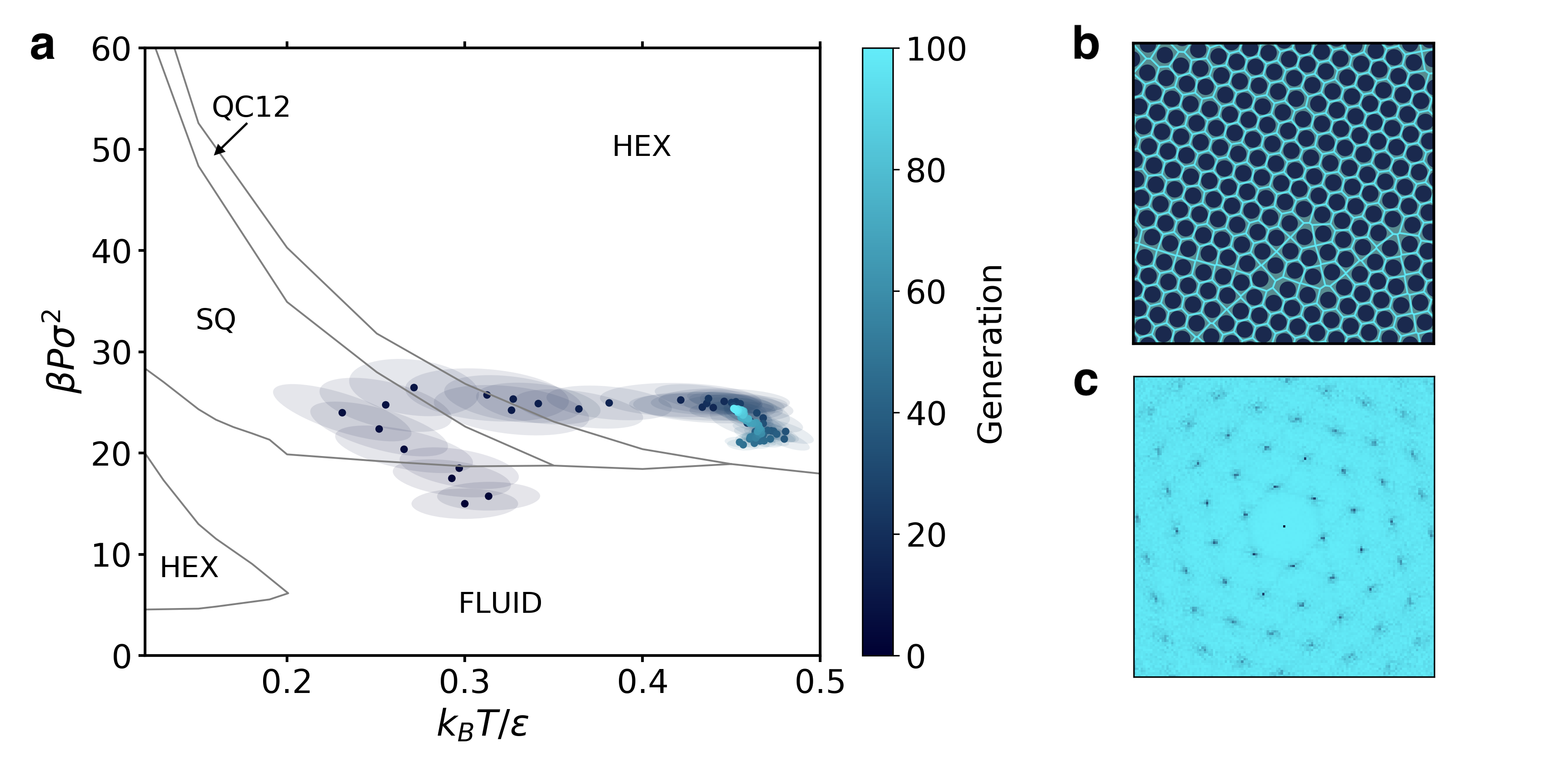}
\caption{{\bf Reverse engineering of the HEX phase in the HCSS model.} (a) Evolution of the Gaussian distribution in the $k_BT/\epsilon-\beta P\sigma^2$ plane. Points and ellipses represent the mean and the covariance matrix (within one standard deviation) of the distribution. The phase diagram in the background is adapted from Ref. \onlinecite{pattabhiraman2015stability}. (b) Representative snapshot of the HEX crystal obtained during the last generation, and (c) its diffraction pattern.}
\label{fig:hex}
\end{figure*}


\bibliography{ref}